\documentclass[aps,prc,showpacs,twocolumn,amssymb,floatfix]{revtex4}
\usepackage{graphicx}
\newcommand{\be}{\begin{equation}}
\newcommand{\ee}{\end{equation}}
\newcommand{\bea}{\begin{eqnarray}}
\newcommand{\eea}{\end{eqnarray}}

\newcommand{\bfsigma}{\mbox{\boldmath $\sigma$}}

\begin{document}
\title{Benchmarking nuclear models for Gamow-Teller response}
\author{E. Litvinova}
\affiliation {Department of Physics, Western Michigan University,
Kalamazoo, MI 49008-5252, USA} \affiliation{National Superconducting
Cyclotron Laboratory, Michigan State University, East Lansing, MI
48824-1321, USA}
\author{B.A. Brown}
\affiliation{Department of Physics and Astronomy, Michigan State
University, East Lansing, MI 48824-1321, USA}
\affiliation{National Superconducting Cyclotron Laboratory, Michigan
State University, East Lansing, MI 48824-1321, USA}
\author{D.-L. Fang}
\affiliation{National Superconducting Cyclotron Laboratory, Michigan
State University, East Lansing, MI 48824-1321, USA}
\affiliation{Joint Institute for Nuclear Astrophysics, Michigan
State University, East Lansing, MI 48824-1321, USA}
\author{T. Marketin}
\affiliation{Physics Department, Faculty of Science, University of
Zagreb, Croatia}
\author{R.G.T. Zegers}
\affiliation{Department of Physics and Astronomy, Michigan State
University, East Lansing, MI 48824-1321, USA}
\affiliation{National Superconducting Cyclotron Laboratory, Michigan
State University, East Lansing, MI 48824-1321, USA}
\affiliation{Joint Institute for Nuclear Astrophysics, Michigan
State University, East Lansing, MI 48824-1321, USA}
\date{\today}

\begin{abstract}
A comparative study of the nuclear Gamow-Teller response (GTR)
within conceptually different state-of-the-art approaches is
presented. Three nuclear microscopic models are considered: (i) the
recently developed charge-exchange relativistic time blocking
approximation (RTBA) based on the covariant density functional
theory, (ii) the shell model (SM) with an extended ``jj77" model
space and (iii) the non-relativistic quasiparticle random-phase
approximation (QRPA) with a Brueckner G-matrix effective
interaction. We study the physics cases where two or all three of
these models can be applied. The Gamow-Teller response functions are
calculated for $^{208}$Pb, $^{132}$Sn and $^{78}$Ni within both RTBA
and QRPA. The strengths obtained for $^{208}$Pb are compared to data
that enables a firm model benchmarking. For the nucleus $^{132}$Sn,
also SM calculations are performed within the model space truncated
at the level of a particle-hole (ph) coupled to vibration
configurations. This allows a consistent comparison to the RTBA
where ph$\otimes$phonon coupling is responsible for the spreading
width and considerable quenching of the GTR. Differences between the
models and perspectives of their future developments are discussed.
\end{abstract}


\maketitle

\section{Introduction}

%In spite of a long and successful history of the field, nowadays the
%low-energy nuclear physics has come to the understanding that a
%reliable high-precision theory of nuclear structure properties is
%still a big challenge.

In the last decades, nuclear physics has greatly expanded its domain
by taking into consideration nuclei away from the valley of
stability that are formed as intermediates in astrophysical
processes leading to synthesis of heavy elements \cite{AGT.07}.
However, in spite of many advances made over decades of research, a
global high-precision theory for the description of structure
properties of these nuclei is still lacking. While nucleosynthesis
studies have strongly benefited from the advances in nuclear
structure models, astrophysical modeling is still suffering from
ambiguities arising from the nuclear physics input. In order to meet
the astrophysical needs, theoretical models should be as microscopic
and universal as possible. In the context of the astrophysical
modeling, it is highly desirable to come to a high-precision
solution of the nuclear many-body problem to enable computation of
masses, matter and charge distributions, spectra, decay and various
reaction rates consistently within the same framework at zero and
finite temperatures.

Although lately the three major concepts in low-energy nuclear
theory have advanced, namely (i) ab-initio approaches, (ii)
configuration interaction models (known also as shell-models) and
(iii) density functional theories (DFT), they still have to be
further developed to satisfy the above mentioned requirements.
Furthermore, each of them has limitations to their applicability
\cite{Bertsch}.
%Figure \ref{nlandscape} shows the chart of the
%nuclei in which the regions where the different approaches can be
%applied are indicated.
%The areas of the nuclear landscape which can be described
%by each of the three theoretical concepts are outlined (in the
%present work we focus on the spin-isospin nuclear properties, e.g.,
%Gamow-Teller response).

The sectors of the nuclear landscape where the applicability
of the different models overlap  are of particular interest because within
these sectors the models can be compared and possibly be used to
constrain each other. Here we focus on the description of
the Gamow-Teller response.
Ab-initio models can replace the
phenomenological input which is traditionally used in the shell
model (SM) with the microscopic effective interaction computed from
the first principles \cite{NQSB.09,CCGIK.09}. In turn, the shell
model with its very advanced configuration interaction concept can
guide the DFT-based developments beyond its standard mean-field and
random phase approximations \cite{NCBBM.12,MLVR.12}. As a feedback,
the extended DFT can provide the SM with single-particle
input for the systems where experimental information is not
available. Thus, in contrast to considering different models as
independently developing alternatives, we rather admit their
complementarity which can be used for their further advancements.

%
%\begin{figure}[ptb]
%\begin{center}
%\vspace{-3cm} \hspace{-0.93cm}
%\includegraphics*[scale=0.34]{972_gt-02.eps}
%\vspace{-1.9cm}
%\end{center}
%\begin{center}
%\vspace{-3.2cm} \hspace{-0.93cm}
%\includegraphics*[scale=0.85]{fig3b.eps}
%\vspace{-0.8cm}
%\end{center}
%\caption{Chart of the nuclei
%\cite{Bertsch}
%representing stable
%nuclei and nuclei found in nature (black), those produced and
%investigated in the laboratory (green) and theoretical limits of
%bound nuclei (yellow). The domain of ab-initio models is the
%lightest nuclei (blue outline), the configuration interaction
%approach is applicable in the pink areas and density functional
%theory covers the region outlined in green.}
%\label{nlandscape}%
%\vspace{-0.2 cm}
%\end{figure}
%

The spin-isospin response is one of the most important properties of
nuclei. The Gamow-Teller (GT) strength distribution, associated with
a spin-transfer $\Delta S$ of one unit, an isospin transfer $\Delta
T$ of one unit, and no angular momentum transfer $\Delta L=0$,
provides information for nuclear beta-decay and other weak processes
in stars. Because GT transitions play an important role in such a
wide variety of astrophysical processes, accurate information is
required for a large fraction of the nuclear chart.
%displayed in Fig. \ref{nlandscape}.
The shell model has been used very
successfully to describe the GT response for nuclei in the $p$, $sd$
and $pf$ shell \cite{bab2001,CMNPZ.05}. The major advantage of these
calculations is that the configuration interaction (CI) method
provides realistic many-body wave functions starting from a
realistic nucleon-nucleon interaction that are complete with regard
to a valence space consisting of a few orbitals near the
Fermi-surface. On the other hand, the major drawback of the SM is
that even a modest increase in the size of the valence space used in
the calculations results in a exponential growth of the CI
dimensions. Nevertheless, recent progress in computer technology,
numerical algorithms, and improved nucleon-nucleon effective
interactions make it possible to overcome these technical
difficulties for certain areas in the chart of nuclei. For example,
a recent shell-model analysis was able to take into account all
relevant nuclear orbitals necessary for a good description of the GT
strength and double beta decay of $^{136}$Xe without recourse to
artificially small quenching factors \cite{hb13}.

Density functional theory is the only candidate that can provide a
description of GTR for the major part of the nuclear chart. However,
until recently, the self-consistent DFT-based studies of the GTR
were confined only by the quasiparticle random phase approximation
(QRPA) \cite{BTF.90,EBDNS.99,PNVR.04,SMEC.98}. Another version of
QRPA employs realistic residual interaction of Brueckner G-matrix
derived from the CD-Bonn nucleon-nucleon potential
\cite{STF88,YRFS09}. This approach has been successfully applied to
description of GTR as well as two-neutrino and neutrinoless double
beta decay. Recently, large-scale calculations for beta-decay
properties of spherical nuclei along the r-process path
\cite{FBS2.13} and neutron-rich deformed nuclei \cite{FBS1.13} have
been reported.

Fragmentation of the GTR has also been extensively addressed, for
instance, within the Quasiparticle-Phonon Model \cite{KS.84} and
second RPA \cite{DNSW.90} (see also references therein), however,
these developments did not aim at a self-consistent description of
GTR and involved adjustable phenomenological effective interactions.
An attempt to describe GTR in medium-mass nuclei within a
self-consistent particle-phonon coupling model based on various
standard Skyrme parameterizations of the density functional have
been reported recently \cite{NCBBM.12}. Earlier, the relativistic
time blocking approximation (RTBA) with fully self-consistent
treatment of particle-phonon coupling based on the covariant DFT
(CDFT) has been developed for the charge-exchange channel. However,
the first application of the charge-exchange RTBA was performed for
the analysis of the spin-dipole strength \cite{MLVR.12}.
%bab ref here

In this article, the Gamow-Teller response of doubly-magic nuclei is
calculated within the frameworks of RTBA, QRPA and SM. We consider
the Gamow-Teller response in the following three doubly-magic
nuclei: (i) $^{208}$Pb, where recently experimental data have become
available up to high excitation energy \cite{WOD.12}, (ii) neutron
rich $^{132}$Sn, which is of the special interest because it
represents a case where the shell-model calculations for GTR are
feasible and have been carried out up to high excitation energies
%even for this heavy nucleus
%so that all the three types of models can be tested against each
%other,
and (iii) $^{78}$Ni.  $^{78}$Ni and $^{132}$Sn play an
important role in some astrophysical r-process scenarios and
influence the r-process abundance distributions for nuclei around N=50 and N=82.
%Predictions of their dynamical properties are of
%great importance.

\section{Microscopic models for spin-isospin response}

\subsection{Relativistic time blocking approximation}
\label{rtba}
Density functional theory can, in principle, provide a
description of the low-energy dynamics for the major part of the
nuclear chart except the lightest nuclei.
%Driven by fast progressing
%disciplines like astrophysics, exotic nuclear structure experimental
%studies and synthesis of superheavy elements, the nuclear DFT has
%achieved a level of sophistication which permits a description of a
%wide range of properties for arbitrarily heavy nuclei including
%those at neutron and proton drip lines.
However, the DFT alone does not allow a high-precision description
of nuclear properties due to very limited treatment of many-body
correlations,
which are especially important for exotic systems at
extremes of nuclear stability. The delicate interplay of various kinds of
correlations is responsible for the binding energy, low-energy spectra,
shapes and decay properties of loosely-bound systems.
Extended DFT is one of the most promising microscopic theories for
providing a consistent input for astrophysical modeling.

Recent extensions of the DFT use the relativistic framework
\cite{Ring.96,VALR.05} and include temporal and spatial
non-localities in the nucleonic self-energies. In medium-mass and
heavy nuclei, the non-local parts of the nucleonic self-energies
modeled in terms of coupling between single-particle and collective
degrees of freedom are treated perturbatively by means of the
nuclear field theory technique \cite{BBBL.77}. The covariant density
functional theory provides a good first approximation to the static
part of the nucleonic self-energy, and a very convenient working
basis for the consistent treatment of its time-dependent non-local
terms \cite{LR.06,LA.11,L.12}. The nuclear response function,
derived consistently within this formalism in the relativistic
time-blocking approximation, involves an energy-dependent residual
interaction which is responsible for the spreading mechanism of
nuclear excitations in both neutral \cite{LRT.08,LRT.10} and
charge-exchange \cite{MLVR.12} channels. No additional adjustable
parameters are introduced within this approach and the few
parameters (8-10) of the CDFT, adjusted at the initial stage to
masses and radii of several characteristic nuclei, remain unchanged.
Further development of the CDFT is proceeding in two directions: (i)
additions beyond the level of the mean-field and random phase
approximations for the description of the ground and excited states,
respectively,
%bab OK?
by inclusion of two-particle two-hole and higher configurations,
%more complex than one-particle-one-hole (1p-1h) configurations,
and (ii) an attempt to provide a microscopic derivation of the
density functional \cite{RVCRS.11}. These two directions are not
independent: only after the proper inclusion of the correlations a
correct comparison to data is possible, that, in turn, gives
conclusions about the origin of the underlying functional.

The RTBA calculations for the GTR are performed in the following
three steps: (i) a relativistic mean field (RMF) solution is
obtained by minimization of the covariant density functional with
NL3 parametrization \cite{NL3}, (ii) phonon spectrum and coupling
vertices for the phonons with $J^{\pi} = 2^+, 3^-, 4^+, 5^-, 6^+$
are obtained by the self-consistent relativistic RPA (RRPA)
solutions \cite{RMGVWC.01} and (iii) the Bethe-Salpeter equation is
solved for the proton-neutron response function with $J^{\pi} =
1^{+}$:
\begin{equation}
R(\omega) = {\tilde R}^{0}(\omega) + {\tilde R}^{0}(\omega)
W(\omega)R(\omega),
\end{equation}
where ${\tilde R}^{0}(\omega)$ is the propagator of the two
uncorrelated quasiparticles in the static mean field and the second
integral part contains the in-medium nucleon-nucleon interaction
$W(\omega)$. The two-body interaction $W(\omega)$ consists of the
following static terms and of the terms depending on the frequency
$\omega$:
\begin{equation}
W(\omega) = V_{\rho} + V_{\pi} + V_{\delta\pi} + \Phi(\omega) -
\Phi(0). \label{inter}
\end{equation}
$V_{\rho}$ and $V_{\pi}$ are the finite-range $\rho$-meson and the
$\pi$-meson exchange interactions, respectively. They are derived
from the covariant energy density functional and read
\cite{PNVR.04}:
\begin{eqnarray}
V_{\rho}(1,2) =
g_{\rho}^2{\vec\tau}_1{\vec\tau}_2(\beta\gamma^{\mu})_1(\beta\gamma_{\mu})_2
D_{\rho}({\bf r}_1,{\bf r}_2) \nonumber\\
V_{\pi}(1,2) = -
\Bigl(\frac{f_{\pi}}{m_{\pi}}\Bigr)^{2}{\vec\tau}_1{\vec\tau}_2({\bf\Sigma}_1{\bf\nabla}_1)
({\bf\Sigma}_2{\bf\nabla}_2)D_{\pi}({\bf r}_1,{\bf r}_2),
\end{eqnarray}
where $g_{\rho}$ and $f_{\pi}$ are the coupling strengths,
$D_{\rho}$ and $D_{\pi}$ are the meson propagators and ${\bf\Sigma}$
is the generalized Pauli matrix \cite{PNVR.04}. The Landau-Migdal
term $V_{\delta\pi}$ is the contact part of the nucleon-nucleon
interaction responsible for the short-range repulsion:
\begin{equation}
V_{\delta\pi}(1,2) =
g^{\prime}\Bigl(\frac{f_{\pi}}{m_{\pi}}\Bigr)^2{\vec\tau}_1{\vec\tau}_2{\bf\Sigma}_1{\bf\Sigma}_2
\delta({\bf r}_1 - {\bf r}_2),
\end{equation}
where the parameter $g^{\prime}$ = 0.6 is adjusted to reproduce
experimental data on the excitation energies of the Gamow-Teller
resonance in $^{208}$Pb and kept fixed in the calculations for other
nuclei, relying on the results obtained in Ref. \cite{PNVR.04}
within the relativistic QRPA. The amplitude $\Phi(\omega)$ describes
the coupling of the nucleons to vibrations (phonons) generated by
the coherent nucleonic oscillations. In the time blocking
approximation it has the following operator form:
\begin{equation}
\Phi(\omega) = \sum\limits_{\mu,\eta}g_{\mu}^{(\eta)\dagger}{\tilde
R}^{0(\eta)}(\omega - \eta\omega_{\mu})g_{\mu}^{(\eta)}, \label{phi}
\end{equation}
where the index $\mu$ numerates vibrational modes (phonons) with
frequencies $\omega_{\mu}$ and generalized particle-vibration
coupling (PVC) amplitude matrices $g_{\mu}^{(\eta)}$, and the index
$\eta = \pm 1$ denotes forward and backward components, in full
analogy with the neutral-channel case \cite{LRT.08}. The
energy-dependent effective interaction of Eq. (\ref{phi}) is
responsible for the spreading mechanism caused by the coupling
between the ph and ph$\otimes$phonon configurations. The phonon
space is truncated by the angular momenta of the phonons at $J^{\pi}
= 6^+$ and by their frequencies at 15 MeV. The ph$\otimes$phonon
configurations are included up to 30 MeV of the excitation energy.
The truncation is justified by the subtraction of the term $\Phi(0)$
in Eq. (\ref{inter}). This subtraction removes double counting of
the PVC effects from the residual interaction, guarantees the
stability of the solutions for the response function and provides
faster convergence of the renormalized PVC amplitude
$\Phi(\omega)-\Phi(0)$ with respect to the phonon angular momenta
and frequencies. This technique is discussed in detail in Ref.
\cite{Tse.13}.
%, because the parameters of the density functional
%have been adjusted to experimental data for ground states and,
%therefore, include the particle-phonon correlations in the static
%approximation.

The strength function $S^{P}(\omega)$
\begin{equation}
S^{P}(E,\Delta )=-\frac{1}{\pi} Im \langle P^{\dagger}R(E+i\Delta)P
\rangle,
\label{strf}%
\end{equation}
gives the spectral distribution of the nuclear response for a
particular external field $P$ which is, in the present case,
expressed by the Gamow-Teller lowering operator:
\begin{equation}
P = \sum\limits_{i=1}^{A}\tau_{-}^{(i)}{\bf\Sigma}_i.
\end{equation}
%bab I do not understand this? Is it energy resolution or spreading width?
%bab if it is energy resolution then it is different for every exp - correct?
A finite value of the imaginary part of the energy variable is
usually taken of the order of the experimental energy resolution, to
make a
consistent
%correct
comparison to data.

\subsection{Quasiparticle random phase approximation based on the realistic N-N interaction}

The non-relativistic proton-neutron (pn) Quasiparticle Random Phase
Approximation
%(pn-QRPA)
has been adopted for the Gamow-Teller (GT) response
% for several decades
\cite{EBDNS.99,MR89,BDEN01}, and it gives good predictions
for the GTR strength distributions with the fulfillment of the Ikeda
sum rule \cite{MR89}. The idea of implementing a realistic nuclear
force in QRPA calculations for both spherical and deformed nuclei
has also been proposed
%decades ago
\cite{STF88,YRFS09}. In this
work, we focus on the spherical nuclei and adopt the spherical
version of the QRPA with realistic forces.

The QRPA concept is based on the introduction of the quasiparticle
creation operator:
\begin{eqnarray}
\alpha_{\tau}^\dagger=u_{\tau}c^\dagger_{\tau}+v_{\tau}c_{\tilde{\tau}},
\end{eqnarray}
where $\tau$ indicates proton or neutron and $c^{\dagger}$ and $c$
are single particle creation and annihilation operators,
respectively. The symbol ``tilde" marks the time-reversed states.
Using these operators with $u_{\tau}$ and $v_{\tau}$ amplitudes of
the nuclear BCS solution, we can construct the pn-excitation phonon
operator in the form:
\begin{eqnarray}
Q^{JM\dagger}_{m}=\sum_{pn}(X^{J}_{m;pn}
A^{JM\dagger}_{pn}-Y^{J}_{m;pn} \tilde{A}^{JM}_{pn}),
\end{eqnarray}
where the two-quasiparticle operators are defined as
$A^{JM\dagger}_{pn}=[\alpha_p^\dagger\alpha_n^\dagger]_{JM}=C^{JM}_{j_p
m_p;j_n -m_n}\alpha_p^\dagger\alpha_n^\dagger$. The energies, the
forward $X$ and backward $Y$ amplitudes are the solutions of the
QRPA equations derived by the variation method \cite{RS.80}:
\begin{eqnarray}
\left( \begin{array}{cc}
A & B  \\
B^{\ast} & A^{\ast}  \end{array} \right) \left(\begin{array}{c}X \\
Y\end{array}\right) =\omega\left(\begin{array}{cc}1 & 0 \\ 0 & -1
\end{array}\right) \left(\begin{array}{c}X \\ Y\end{array}\right),
\label{qrpa}
\end{eqnarray}
where $A_{pn,p'n'}=[A_{pn},[H,A^\dagger_{p'n'}]]$ and
$B_{pn,p'n'}=[{A}_{pn}^\dagger,[H,\tilde{A}_{p'n'}^\dagger]]$. The
Hamiltonian and the detailed expressions for $A$ and $B$ matrices
with realistic interactions are presented in Ref.~\cite{STF88}.

From the diagonalization of Eq. (\ref{qrpa}), we can obtain the
eigenvalues $\omega_m$ and eigenvectors $X_m$ and $Y_m$ which are
the energies and the amplitudes of the QRPA excitations. With the
realistic forces, we can determine the energies of the excited
states in odd-odd daughter nucleus with respect to its ground state
(the one with the lowest eigenvalue), and denote: $E_m = \omega_m -
\omega_{g.s.}$, where the index $m$ numerates the solutions of Eq.
(\ref{qrpa}). The matrix element of the GT$^-$ transition can be
written as:
\begin{equation}
M_{m}^{GT^-} = \sum\limits_{pn}\langle
p||\tau_{-}\bfsigma||n\rangle\bigl(u_p v_n X^{1^+}_{m;pn}-v_p u_n
Y^{1^+}_{m;pn}\bigr), \label{mgt}
\end{equation}
and the GT strength function is expressed as follows: \be
S^{GT^-}(E)=\sum_{m} \delta(E-E_m) |M_{m}^{GT^-}|^2. \ee

For the QRPA A-matrices of Eq. (\ref{qrpa}), we use the single
particle energies (spe) obtained from the SkX mean field \cite{SKX}.
%(for the sake of comparison we also used spe of SkX with tensor
%forces \cite{BDOAS06}).
The realistic interaction in the form of G-matrix elements is
obtained from the CD-Bonn potential \cite{Mac00} and adopted here in
both particle-hole (ph) and particle-particle (pp) channels as the
residual interaction for the QRPA. No proton-neutron pp-interaction
is included in the calculations because closed shell nuclei with
relatively large asymmetry between the neutron and proton numbers
will be considered. In general, the neutron-neutron and
proton-proton pairing strength is adjusted to reproduce the observed
pairing gaps by the five-point formula~\cite{AW03}. Thus, for the
residual interaction we have two adjusted parameters for the
particle-hole and particle-particle channels $g_{ph}$ and $g_{pp}$.
The parameter $g_{ph}$ is fitted to reproduce the GTR centroid and
we adopt $g_{ph}=1$ if the experimental data is not available. The
constant $g_{pp}$ is fixed equal to $0.6$ to avoid collapse of the
solution of the QRPA equation. The GTR centroid is not sensitive to
$g_{pp}$.

In this work we consider doubly-magic nuclei, for which the BCS
solutions show that there is a sharp change of occupation
probabilities around the Fermi surface and that QRPA calculations
are mainly reduced to RPA calculations. In the RPA limit the
importance of $g_{pp}$ is, in turn, reduced. However, the strength
of the pairing interaction is kept finite to retain the calculation
scheme established for nearly the entire chart of nuclides.

\subsection{Shell model (SM)}
\label{sm}

One of goals of the large-basis nuclear shell model approach is to
use a complete basis within a limited set of single-particle states
near the Fermi surface. This method is used extensively in light
nuclei such as in the sd-shell (A=16-40)  pf-shell (A=40=80) mass
regions \cite{bab2001}. All of orbitals required to obtain the Ikeda
sum-rule for Gamow-Teller strength are contained in the model space.
One of the observations for these model spaces is that the
experimental B(GT) values extracted from beta decay and
charge-exchange reactions is about a factor of two smaller than
those calculated. Thus, for the $sd$ or $pf$ model spaces one needs
a reduction (quenching) factor for the Gamow-Teller operator of
0.74-0.77 \cite{sdgt,pfgt}. This quenching is consistent with
theoretical calculations of the operator renormalization obtained in
second-order perturbation theory \cite{arima,towner}.

For heavier nuclei the number of basis states even for a modest
number of orbitals grows exponentially as the number of valence
nucleons increases. Thus, the shell model applications are
restricted to semi-magic nuclei or those near double-magic nuclei
such as $^{132}$Sn or $^{208}$Pb. Often the orbitals used in the
model space are not sufficient to accommodate the Ikeda sum rule.
For example, in the region north-west of $^{132}$Sn the ``jj55"
model space if often used. The notation jj55 represents the five
orbitals $0g_{7/2},1d_{5/2},1d_{3/2},1s_{1/2},0h_{11/2}$ in between
the magic numbers 50 and 82 for protons and neutrons. The $0g_{9/2}$
and $0h_{9/2}$ orbitals need to be added to satisfy the Ikeda sum
rule.

One of the applications of these calculations is for the double beta
decay of $^{136}$Xe. Up until recently the jj55 model space has been
used with the understanding that some renormalization of the
operators may be related to the restricted model space. For the
Gamow-Teller operator that enters into the two-neutrino double beta
decay the renormalization might be fixed by reproducing some single
and double beta decay rates. In  \cite{caur-136xe} a quenching
factor of 0.45 was used to obtain the observed two-neutrino rate.
The question is then to what extent other operators such as those
for neutrino-less double beta decay are renormalized.

Recently the jj55 model space was enlarged to jj77 where the
configurations involving $0g_{9/2}$ and $0h_{9/2}$ orbitals were
included that are required to obtain the Ikeda sum rule. The details
for the derivation of the Hamiltonian are described in \cite{hb13}.
In brief, it is obtained with realistic nucleon-nucleon interaction
renormalized to the jj77 model space, and with single-particle
energies adjusted to reproduce the experimental values observed in
$^{131}$Sn and $^{133}$Sb. This is typical of all shell-model
calculations. If single-particle energies are not available from
experiment one must rely on those obtained from
best Skyrme Hartree-Fock or RMF model extrapolations.

In this work the shell-model calculations for the GT response of
$^{132}$Sn are performed. Two truncations were used. The simplest
called TDA has a closed-shell configuration for $^{131}$Sn in the
jj55 model space with the addition of two one-particle one-hole
final-state configurations, $0g_{9/2}^{(-1)}-0g_{7/2}^{(1)}$ and
$0h_{11/2}^{(-1)}-0h_{9/2}^{(1)}$. The GT distribution for this
is very similar to that obtained with QRPA.
For second called TDA + (1p-1h),
these TDA configurations were coupled to 1p-1h ``vibrations" of the
$^{132}$Sn core that are obtained within the jj77 model space.

%When I get back around July 10th
%I will make a figure for the simple jj55 model
%space where the Ikeda sum-rule is not satisfied.
%
\section{Gamow-Teller strength in doubly-magic nuclei}

%\subsection{$^{208}$Pb}
%
\begin{figure}[ptb]
\begin{center}
\vspace{-5mm}
\includegraphics[scale=0.45]{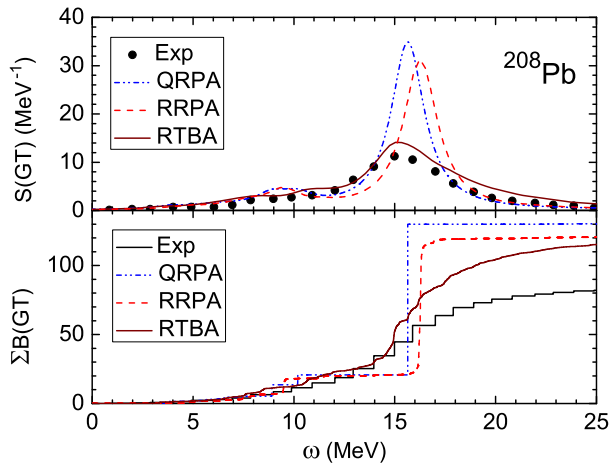}\\
\end{center}
\vspace{-0.5cm} \caption{The theoretical and experimental
Gamow-Teller strength distributions in $^{208}$Pb (upper panel) and
their cumulative sums (lower panel).}
\label{pbgtexp}%
\vspace{-0.5cm}
\end{figure}
%
%In the QRPA calculations for $^{208}$Pb, the particle-hole strength
%is adjusted to the experimental centroid of the GTR at about 16.0
%MeV, so that $g_{ph} = 1.1$.
In Figure \ref{pbgtexp}, we show the results for the GTR in
$^{208}$Pb obtained within the QRPA, RRPA and RTBA, compared to data
of Ref. \cite{WOD.12}. The non-relativistic QRPA results are folded
by the Lorentz distribution with one MeV width which is close to the
energy resolution of the experiment. The parameter $g_{ph} = 1.15$
is adjusted to reproduce the GTR centroid.
%bab I discussed with the Dong-Liang - it should left out - he will change the figure.
%The QRPA here is based on the three
%different mean-field parameterizations: the SkX without tensor
%force, and SkX with different tensor forces TFa and TFb
%\cite{BDOAS06} and the parameter $g_{ph} = 1.15$ is adjusted to
%reproduce the GTR centroid for the SkX without tensor.
%bab where is the data in the bottom panel?
%We see that the inclusion of tensor
%forces in the Skyrme mean-field has different impacts on the GT
%distribution for the doubly magic nucleus $^{208}$Pb: the TFa shifts
%the position of the GTR by about 1 MeV and weakens the low-lying
%structure at $E_{ex}\sim$ 10 MeV. Another tensor force TFb does not
%influence the GTR strength.
%This confirms the conclusion of Ref.
%\cite{BDOAS06}, where the TFb is suggested as a better
%parametrization than the TFa force.
%The Ikeda Sum rule is strictly fulfilled in the presented realistic
%QRPA calculations within the
The QRPA model space, including pn-configurations up to ~45 MeV,
accommodates the exact Ikeda sum rule while 3\% of the total
$B(GT_{-})$ is beyond the considered 25 MeV energy interval and the
total $B(GT_{+})$ is equal to 0.21. Without introducing quenching
factors in front of the calculated strength function the
experimentally observed total strength \cite{WOD.12} is by factor
0.62 smaller than that obtained in the QRPA.
% , and that
%bab added
%the QRPA cannot account for the observed spreading width.

The GTR within the relativistic approaches RRPA and RTBA described
in Section \ref{rtba} has been calculated using the smearing
parameter $\Delta$ = 1 MeV. The RRPA calculations, neglecting the
last two terms of Eq. (\ref{inter}), produce a strength distribution
which is very similar to the non-relativistic QRPA calculations with
the major peak at 16.5 MeV and a low-energy peak structure around 10
MeV. The exact Ikeda sum rule is accommodated within the model space
of pn-configurations between -1800 MeV and 100 MeV, so that ~8\% of
the $B(GT_-)$ is at large negative energies because of the
transitions to the Dirac sea \cite{PNVR.04}. While both QRPA and
RRPA do not account for spreading effects, within RTBA the GTR
acquires the spreading width because of the coupling between the ph
and ph$\otimes$phonon configurations, so that the additional 5\% of
the sum rule goes above the considered energy region, while the
total $B(GT_{+})$ is equal to 0.34. Comparison to data shows that
the spreading effects which are taken into account in the RTBA are
reproduced very well.

\begin{figure}[ptb]
\begin{center}
%\vspace{-5mm}
\includegraphics[scale=0.42]{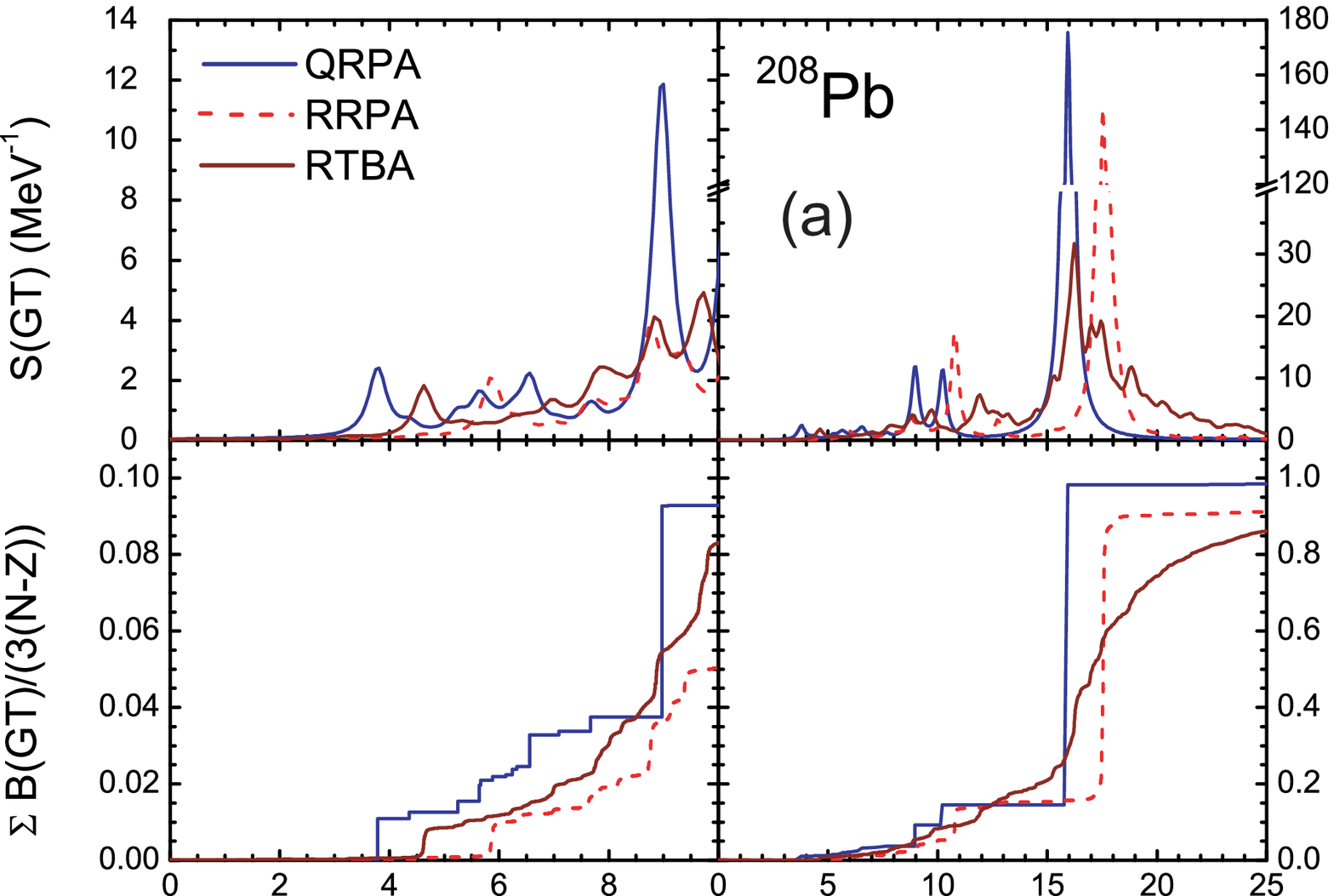}\\
\vspace{3mm}
\includegraphics[scale=0.42]{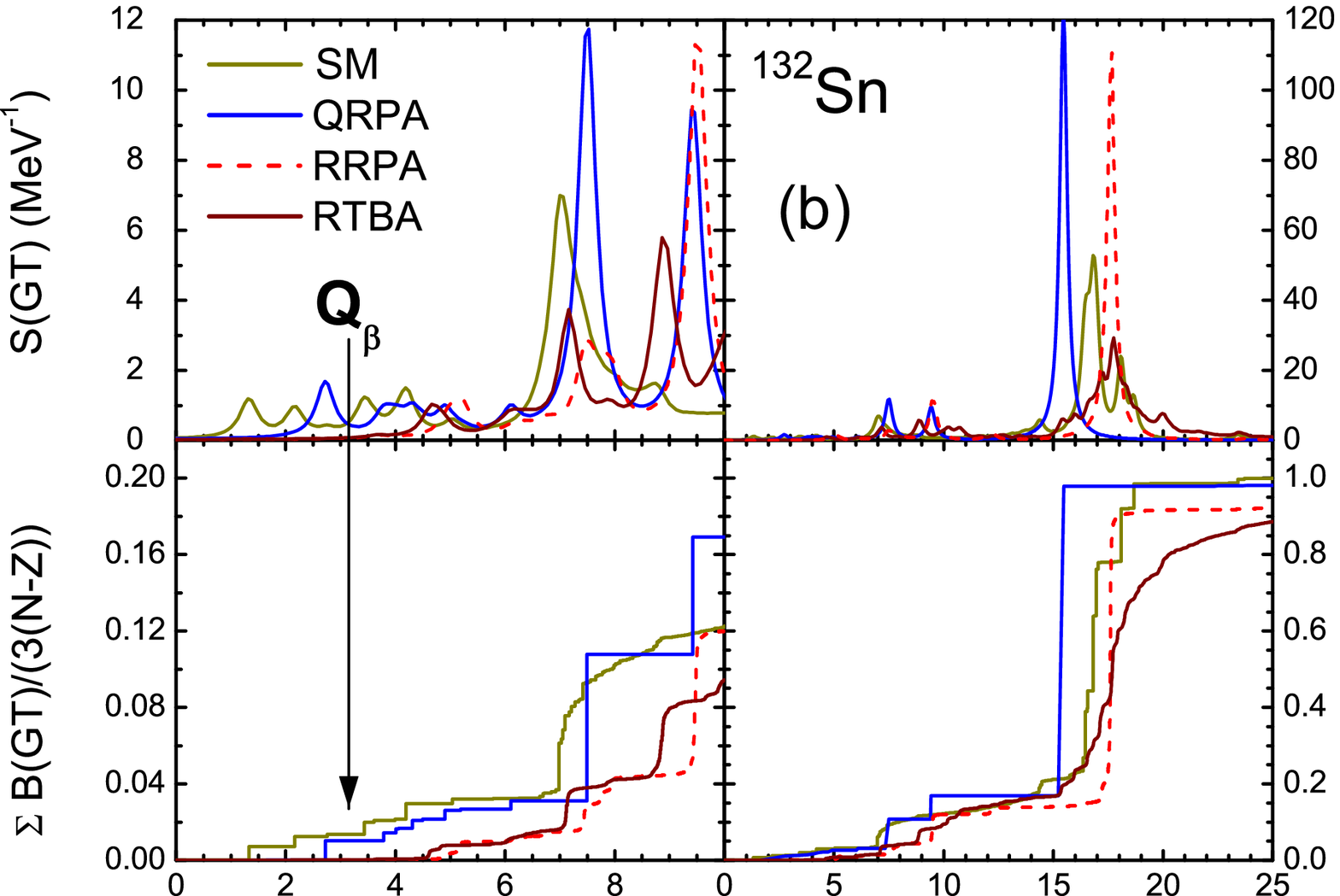}\\
\vspace{3mm}
\includegraphics[scale=0.42]{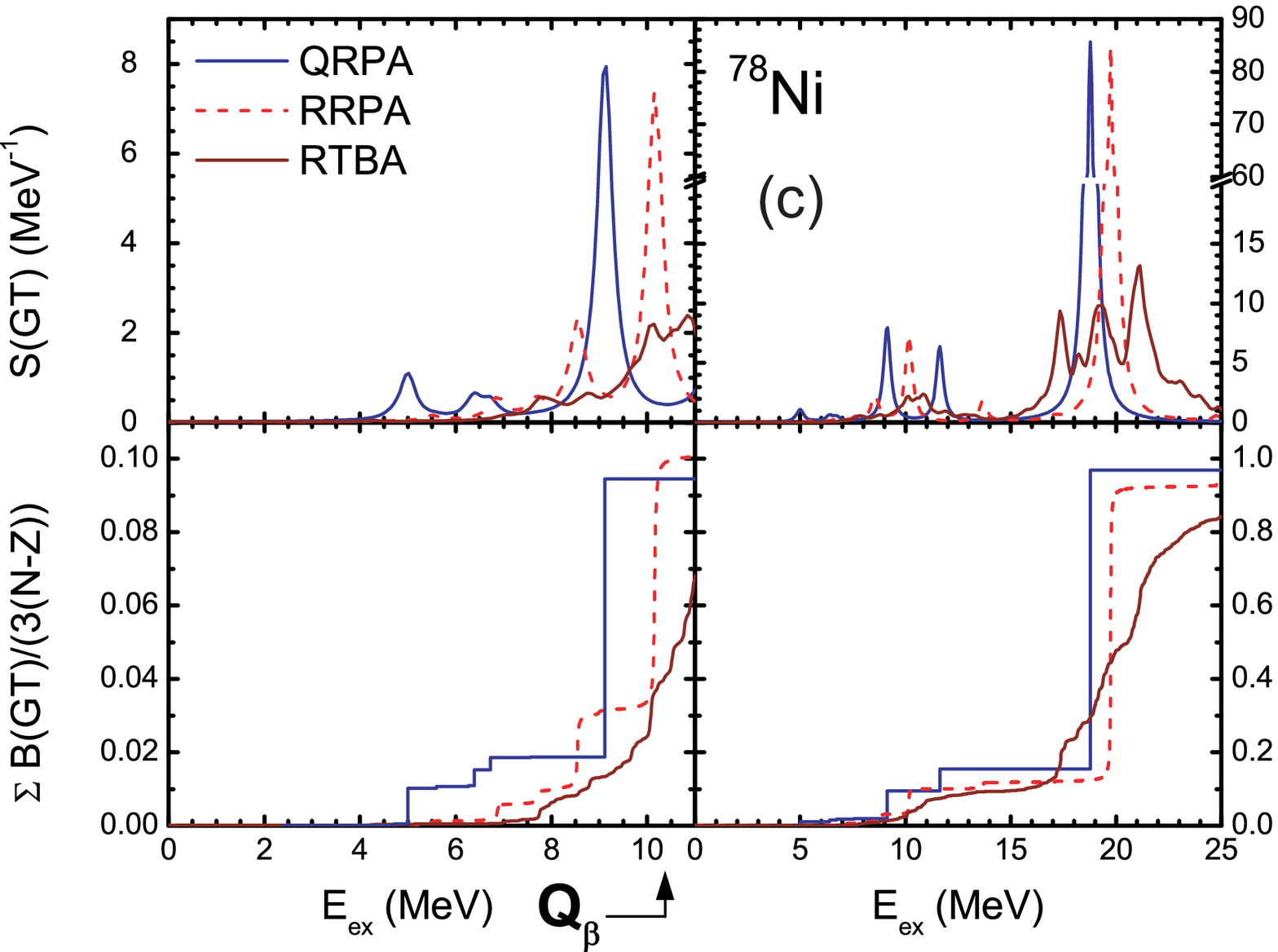}
\end{center}
\vspace{-0.5cm} \caption{The Gamow-Teller strength distribution in
$^{208}$Pb (a) and in neutron-rich $^{132}$Sn (b) and $^{78}$Ni
(c). The SM curves for $^{132}$Sn are from the shell-model
results for TDA coupled to 1p-1h ``vibrations".}
\label{gt}%
\vspace{-0.5cm}
\end{figure}

A more detailed analysis of the non-relativistic and relativistic
calculations for the GTR in $^{208}$Pb has been presented in Fig.
\ref{gt}(a) for both the overall GTR structure (right panels) and
the low-lying part(left panels). Compared to Fig. \ref{pbgtexp}, we
have reduced the smearing parameter $\Delta$ to 200 keV, to see more
detailed features of the GTR. Besides this,
%we
%also have different horizontal axis than that in Fig. \ref{pbgtexp}
%where $\omega$ is the energy transfer or, in the other words, the
%energy of the excited state in the daughter nucleus relative to the
%ground states of the parent nucleus.
in Fig. \ref{gt} we show the calculated spectra
%%with respect to the excitation energies
relative to the ground states of daughter nuclei. Since in the
present version of the QRPA the effective interaction is not related
to any self-consistent mean field, the ground state energies are not
defined in this model. However, the single-particle energies
entering the QRPA equations are adjusted to data, therefore, for the
QRPA we find consistent to use the experimental $Q_{\beta}$ values.
In contrast, for the self-consistent RRPA and RTBA, in which the
effective interaction is the exact second variational derivative of
the covariant energy density functional with respect to the density
matrix, we use the following formula: $Q_{\beta} = M(Z,N) -
M(Z+1,N-1)$. Here $M(Z,N)$ and $M(Z+1,N-1)$ are the masses of the
mother and the daughter nuclei, respectively, calculated in the
relativistic mean field by the minimization of the CEDF. Thus, for
$^{208}$Pb the main GTR peak appears in the RRPA at about 1.5 MeV
higher than in the QRPA. When the coupling to the ph$\otimes$phonon
configurations is included by the RTBA, the major GTR peak shifts
down by the same 1.5 MeV, however, the centroid remains at the same
energy as in RRPA. For the low-lying part of the strength
distribution, in particular, for the first excited state, the RTBA
calculation shows a much better agreement with the QRPA than the
RRPA result. Correlations of the PVC type in the RTBA increase the
nucleon effective mass and single-particle level density up to their
realistic values \cite{LR.06}, which, in turn, causes spreading of
the strength to the low-energy region. Single-particle levels used
in the QRPA are adjusted to data and, therefore, account for the
self-energy part of these correlations implicitly. An additional
fine tuning of the $g_{ph}$ and $g_{pp}$ parameters takes into
account effectively the phonon-exchange PVC correlations, so that
the QRPA built in this way describes successfully gross features of
the excitation spectra without explicit treatment of the
correlations beyond the one-phonon ones.
%, very reasonably.
%\subsection{$^{132}$Sn}
%

The GTR for $^{132}$Sn is shown in fig. \ref{gt}(b) in the same
fashion as for $^{208}$Pb. The same parameter sets as for $^{208}$Pb
are used here within RRPA and RTBA. We have included also results
from the shell model calculations for TDA coupled to 1p-1h
``vibrations" described in Section \ref{sm}. For the QRPA we used
the bare value of $g_{ph}=1$ without a renormalization, since
experimental data on the GT transition strength are not yet
available for this nucleus.
%%%%%%%%%%%%%% to Referee # 2
Here we can see how the QRPA results will compare with the other
models without renormalization, because setting $g_{ph} = 1$ is the
usual practice for experimentally unknown nuclei.
%%%%%%%%%%%%%
The shell model TDA results (not shown in Fig. 3) are very similar
to the QRPA results. The gross structures of the GTR obtained within
QRPA and RRPA are very similar except the fact that the overall RRPA
strength distribution is shifted upwards by $\approx$~2 MeV relative
to the QRPA strength. This difference appears when we relate the GTR
strength to the ground state of the daughter nucleus, which is done
self-consistently in the RRPA and using the experimental $Q_{\beta}$
values in QRPA. The inclusion of coupling to the ph$\otimes$phonon
configurations by the RTBA leads to a strong fragmentation of the
major GTR peak. However, the centroid of the distribution does not
shift, and the low-lying strength distribution changes very little.

%To gain a better understanding of the differences observed between
%the results of the relativistic and non-relativistic methods for
%$^{132}$Sn, the GTR strength distribution calculated within the
%shell-model is added to Fig. \ref{gt}(b).
Within the SM, coupling between the Tamm-Dancoff proton-neutron
phonons and particle-hole core vibrations produces strength which is
more fragmented than seen in the QRPA and RRPA calculations, but
less than in the RTBA calculations, because of the truncation of the
SM valence space.
%%%%%%%%%%%%%%%%%% In response to Referee # 2
The SM strength is also related to the ground state of the daughter
nucleus. As the single-particle energies in the SM are adjusted to
data and the ground state energies are not defined in this model, it
is consistent to use the experimental $Q_{\beta}$ values. This
practically means that the energy of the first $1^+$ state in the SM
matches its experimental position.
%To have a realistic description of the low-lying part of the GTR
%which is closely related to the $\beta$-decay half-lives, the energy
%of the first $1^+$ state is set to match the experimental value.
%bab where is the data shown?
%%%%%%%%%%%%%%%%%%
Thus, we see that QRPA overestimates the experimental $1^+_1$ energy
by 1.5 MeV, but underestimates the centroid predicted by the SM
calculation by a similar value. The latter, however, can be changed
by tuning the $g_{ph}$ parameter.

Below seven MeV, there is basically the same amount of strength for
the SM and QRPA, while in the Q-value window, due to over-predicted
$1^+$ energy, less strength contributes to $\beta$-decay for QRPA.
As for the RTBA, while the GTR centroid and width are in a
reasonable agreement with the SM calculations, the spreading to the
low-energy region is weaker. However, there are at least two
mechanisms which are not taken into account in the amplitude
$\Phi(\omega)$ of Eq. (\ref{phi}) in the present calculations:
coupling to pairing vibrations and
%backwards going ('zigzag') diagrams
the ground state correlations caused by the ph$\otimes$phonon
configurations in the response function. These two effects, along
with higher-order particle-vibration couplings, can reinforce the
spreading to the low-energy region and will be included on the next
step of the charge-exchange RTBA development. All the models exhaust
the Ikeda sum rule completely within their model spaces, but it can
be seen from Fig. \ref{gt}(b) that
 2\%, 8\% and 12\% of the sum rule is beyond the considered energy region
in the QRPA, RRPA and RTBA calculations, respectively.
%
%%%%%%%%%%%%%%%%%%%%%%% In response to Referee # 2
\begin{table*}[ptb]
\caption{Reduced matrix elements for the neutron-proton transitions
which contribute mostly to the strength of the major GT peak (GTR)
and to the strongest peak at low energy, computed for the considered
models. See text for detailed explanations.} \label{table}
\begin{center}
%\vspace{6mm}
\tabcolsep=0.5em \renewcommand{\arraystretch}{1.0}%
\begin{tabular}
[c]{cccccccccc}\hline\hline
 &  & \multicolumn{2}{c}{QRPA} & \multicolumn{2}{c}{RRPA} & \multicolumn{2}{c}{RTBA} &
 \multicolumn{2}{c}{SM
 }\\
\hline
 $m$& pn& $\langle$p$\parallel\tau_-\bfsigma\parallel n\rangle$ & $X_{m;pn}^{1+}$ &
 $\langle$p$\parallel\tau_-\bfsigma\parallel n\rangle$ & $X_{m;pn}^{1+}$ &
 $\langle$p$\parallel\tau_-\bfsigma\parallel n\rangle$ & $X_{m;pn}^{1+}$ &
 $\langle$p$\parallel\tau_-\bfsigma\parallel n\rangle$ & $X_{m;pn}^{1+}$ \\
\hline\hline
 & $\pi h_{9/2}-\nu h_{11/2}$ &4.67 & 0.74 & 4.54 & 0.68 & 4.54 & 0.37 & 4.67 & 0.31 \\
%\hline
 & $\pi g_{7/2}-\nu g_{9/2}$ &4.22 & 0.47 & 4.10 & 0.49 & 4.10 & 0.32 & 4.22 & 0.26 \\
%\hline
GTR & $\pi h_{11/2}-\nu h_{11/2}$ &3.77 & 0.23 & 3.60 & 0.22 & 3.60 & 0.12 & 3.77 & 0.12 \\
%\hline
 & $\pi d_{3/2}-\nu d_{5/2}$ &3.10 & 0.22 & 3.01 & 0.21 & 3.01 & 0.11 & 3.10 & 0.13 \\
 & $\pi g_{7/2}-\nu g_{7/2}$ &-2.49 & -0.17 & -2.32 & -0.16 & -2.32 & -0.08 & -2.49 & -0.07 \\
\hline
 & $\pi g_{7/2}-\nu g_{9/2}$ &4.22 & 0.56 & 4.10 & 0.35 & 4.10 & 0.06 & 4.22 & 0.22 \\
%\hline
 & $\pi d_{3/2}-\nu d_{5/2}$ &3.10 & -0.49 & 3.01 & -0.62 & 3.01 & -0.14 & 3.10 &  \\
Low- & $\pi h_{11/2}-\nu h_{11/2}$ &3.77 & -0.33 & 3.60 & -0.15 & 3.60 & -0.15 & 3.77 & -0.27 \\
%\hline
energy & $\pi d_{5/2}-\nu d_{5/2}$ & 2.90 & -0.31 & 2.77 & -0.30 & 2.77 & -0.10 & 2.90 & -0.17 \\
peak & $\pi s_{1/2}-\nu s_{1/2}$ & 2.45 & -0.31 & 2.36 & -0.34 & 2.36 & -0.12 & 2.45 & -0.16 \\
 & $\pi h_{9/2}-\nu h_{11/2}$ & 4.67 &  & 4.54 & 0.21 & 4.54 & 0.01 & 4.67 & 0.14 \\
\hline\hline
\end{tabular}
\end{center}
\vspace{-7mm}
\end{table*}

In order to compare the models on a deeper level, we have considered
the contributions of the individual neutron-proton transitions to
the strength distributions, namely to the main GT peak and to the
most pronounced low-lying peak. Table \ref{table} shows these
contributions in terms of the reduced matrix elements. For all the
models, the matrix elements of the transition densities entering Eq.
(\ref{mgt}) can be uniformly defined as:
\begin{equation}
X_{m,pn}^{1+} = \langle m\parallel[c_p^{\dagger}c_n]^{1+}\parallel
g.s.\rangle ,
\end{equation}
where the nucleonic creation and annihilation operators
$c_p^{\dagger}, c_n$ are related to the respective basis states. The
corresponding backwards going amplitudes $Y_{m,pn}^{1+}$ vanish
identically in RRPA, RTBA and SM, and their contributions in QRPA
are negligible. As mentioned above, the occupation numbers $u_p$ and
$v_n$ are very close to 0 and 1, respectively, in QRPA, and take
these values exactly in the other three models. Thus, the products
of the reduced matrix elements
$\langle$p$\parallel\tau_-\bfsigma\parallel n\rangle$ and
$X_{m,pn}^{1+}$ displayed in Table \ref{table} contribute to the sum
of Eq. (\ref{mgt}).

For the GTR one can see that the five largest contributions are
represented by the same neutron-proton transitions in all four
models and these contributions are coherent. The absolute values of
the matrix elements are very close for QRPA and RRPA. Both RTBA and
shell model show some reduction of the transition densities,
compared to the simpler models, because of the fragmentation
effects, but the absolute values of the amplitudes $X_{m,pn}^{1+}$
are close to each other. The matrix elements
$\langle$p$\parallel\tau_-\bfsigma\parallel n\rangle$ are also very
similar in their absolute values for the relativistic and
non-relativistic models, and their minor differences reflect the
respective differences of the radial wave functions.

The structure of the low-lying GT strength is more sensitive to the
differences in single-particle structure and effective interaction
and shows more variations from model to model which makes an
analysis more difficult. In Table \ref{table} we compare microscopic
structure of the strongest low-energy peak below the GTR. The
position of the peak depends on the model (see Fig. \ref{gt}b), but,
nevertheless, its structure composition reveals considerable
similarities. The first 5-6 major contributions come from the same
neutron-proton transitions, although their numerical values are
different in different models. One can notice, for instance, that
the leading component varies from model to model and in all the
models except RTBA the leading or the next-to-leading component is
out of phase compared to the others which are, in turn, in phase.
Thus, the structure of the low-energy peak reveals some destructive
interference, in contrast to the GTR peak. In the RTBA the
interference is mostly constructive, but the matrix elements
$X_{m,pn}^{1+}$ of the leading components have small absolute values
since they obey the extended normalization condition (see Eq. (63)
of Ref. \cite{LRT.07}), which includes sizable phonon coupling
contribution. In general, the structure of the low-energy peak in
the two models beyond (Q)RPA (RTBA and shell model) shows
similarities as well as differences which are expected to be less
pronounced if correlations of 3p3h nature will be included. This is
the most natural extension of these two models, which is feasible at
the current theoretical and computational capacities.
%%%%%%%%%%%%%%%%%%%%%%%

Another doubly magic nucleus of great importance for astrophysics is
$^{78}$Ni, for which the calculated GTR distributions are displayed
in Fig. \ref{gt}(c).
%%%%%%%%%%%%%%%%%%%%%%% in response to Referee #2
Only three of the considered models: QRPA, RRPA and RTBA are applied
to the GTR in this nucleus. Since shell model calculations are based
on experimental data about single-particle energies which are not
yet known for $^{78}$Ni, shell model calculations are not presented.
%%%%%%%%%%%%%%%%%%%%%%% in response to Referee #2
We keep the self-consistent calculation scheme for RRPA and RTBA and
use $g_{ph} = 1$ without renormalization for QRPA, because of the
absence of experimental data.
%%%%%%%%%%%%%%%%%%
Like in the previous case, very similar distributions for GTR are
obtained within the QRPA and RRPA calculations, except for the
positions of the peak structures. They differ by $\sim$~1~MeV, which
is within the range of reasonable tuning for the $g_{ph}$ parameter
used in QRPA. The RTBA gives a much richer structure in the region
of the main GT peak, and the low-energy fraction is slightly
diminished compared to RRPA. The results for the cumulative sums are
similar to the previous cases: 8\% and 17\% of the RRPA and RTBA
total GT$_-$ strength, respectively, are beyond the considered 0-25
MeV energy region while only 3\% of the total strength is beyond
this interval in QRPA. As before, the Ikeda sum rule is exhausted
within the full model spaces in all three models.

The nucleus $^{78}$Ni is far from the valley of stability and its
$\beta$-decay Q-value is 10.37 MeV, so that relatively many GT
transitions are involved in the $\beta$-decay. Within the QRPA the
integral strength of these transitions is about one tenth of the
total strength, but only the lowest-energy portion contributes to
$\beta$-decay rate because of the phase space factor. While for RRPA
nearly the same amount of strength has been involved, all the
strength has been distributed at higher energies. The RTBA spreads
the strength more widely, some of the strength is shifted to higher
energies, and compared to RRPA, there is a reduction of the decay
width. This can be interpreted as the effective quenching of the
RRPA strength, and with quenching factors defined in this way, one
can obtain an approximate decay width from RRPA or QRPA
calculations.

\section{Conclusions and outlook}

We compare Gamow-Teller response of doubly-magic nuclei computed
within the newly developed proton-neutron relativistic time blocking
approximation based on the CDFT, QRPA with G-matrix effective
interaction and the large-basis shell model.
% advanced in extension of
The QRPA and RTBA models are successfully tested, bench marked to
experimental data on GTR in $^{208}$Pb and applied to predict GTR in
neutron-rich doubly-magic $^{78}$Ni nuclei. All three models are
applied to GTR in $^{132}$Sn nucleus allowing, for the first time, a
comprehensive comparative study.

Such a comparison turns out to be very constructive in defining
strong and weak points of the theory and to determine future
directions. We have demonstrated, by the choice of the appropriate
physics case of the GTR in $^{132}$Sn, that very different
theoretical models can constrain each other. QRPA and SM, based on
the realistic interactions, work well together as the SM complements
%compensates
the deficiency of the configuration mixing in QRPA. The SM helps to
fix the flexible parameters of the QRPA including the explicitly
missed dynamics, that makes the QRPA a useful
%reliable
tool for all nuclei.
%in the considerably larger applicability area.
The RTBA can, to a
%bab
certain extent, provide information that is missing in QRPA and, in addition, can
provide
%bab
part of the quenching factors that are needed for QRPA and SM. In
cases where the RTBA model space includes a sufficiently large
amount of configurations and finite momentum transfer is taken into
account \cite{MMPV.12}, RTBA has a potential to describe the overall
GTR quenching fully microscopically, except the contribution from
the delta-isobar which is found to be small \cite{arima,towner}.

Comparison between the RTBA and SM calculations has become possible
in this work for the first time.
%bab and it is of the special interest.
Spreading effects of the high-energy GTR mode in $^{132}$Sn are
described here within the SM and RTBA on the same level of the
configuration complexity, namely, particle-hole coupled to the core
vibration and rather similar GT strength distributions are obtained.
Thus, SM provides further guidance on inclusion of higher-order
correlations into the RTBA that could improve its performance for
the low-energy region. In turn, RTBA is an advancement in
%bab
partly
resolving
the quenching problem, and, in addition, the relativistic mean field
extended by the particle-vibration coupling \cite{LA.11,L.12} can
provide the SM with the single-particle energies for nuclei where
these energies are not available from data.

Starting from the presented results, further advancements of the
discussed methods are anticipated. Data on the overall GTR
distribution and low-lying strength  in $^{132}$Sn are expected from
future measurements of spin-isospin properties of exotic nuclei at
the rare isotope beam facilities. Such data will provide decisive
arguments to constrain many-body coupling schemes of the RTBA and SM
as well as the underlying nuclear effective interactions.

\bigskip\leftline{\bf ACKNOWLEDGEMENTS}
We acknowledge support from US-NSF grants PHY-1068217, PHY-0822649
(JINA), PHY-1102511 and PHY-1204486. E.L. acknowledges also the
support from NSCL.

\end{document}